\documentclass[journal=cmatex, manscript=article, layout=twocolumn, amsmath,amssymb,nobibnotes,unsortedaddress,superscriptaddress,usetitle=true,dvipdfmx]{achemso}
\setkeys{acs}{usetitle = true}

\usepackage{achemso}
\usepackage{amsmath}
\usepackage{amssymb}
\usepackage{bm}
\usepackage{dcolumn}
\usepackage{graphicx}
\usepackage{mciteplus}
\usepackage{tabularx}
\usepackage{color}
\usepackage{url}
\usepackage{threeparttable}
\usepackage[numbers]{natbib}

\SectionNumbersOff
\title{Stabilization mechanism of tetragonal structure in 
hydrothermal synthesized BaTiO$_3$ nanocrystal}
\author{Kenta Hongo}
\email{kenta_hongo@mac.com}
\affiliation{Research Center for Advanced Computing Infrastructure,
  JAIST, Asahidai 1-1, Nomi, Ishikawa 923-1292, Japan}
\altaffiliation{National Institute for Materials Science (NIMS), 
1-2-1 Sengen, Tsukuba, Ibaraki 305-0047, Japan}
\altaffiliation{Computational Engineering Applications Unit, RIKEN, 
2-1 Hirosawa, Wako, Saitama 351-0198, Japan}
\altaffiliation{PRESTO, JST, 
4-1-8 Honcho, Kawaguchi, Saitama 332-0012, Japan}
\author{Sinji Kurata}
\affiliation{Department of Applied Chemistry, Faculty of Engineering, 
Kyushu University, Fukuoka 819-0395, Japan}
\author{Apichai Jomphoak}
\affiliation{National Electronics and Computer Technology Center (NECTEC),
112 Phahon Yothin, Klong Luang, Pathumthani 12120, Thailand}
\author{Miki Inada}
\affiliation{Center of Advanced Instrumental Analysis, Kyushu University}
\author{Katsuro Hayashi}
\affiliation{Department of Applied Chemistry, Faculty of Engineering, 
Kyushu University, Fukuoka 819-0395, Japan}
\author{Ryo Maezono}
\affiliation{School of Information Science,
  JAIST, Asahidai 1-1, Nomi, Ishikawa 923-1292, Japan}
\altaffiliation{Computational Engineering Applications Unit, RIKEN, 
2-1 Hirosawa, Wako, Saitama 351-0198, Japan}
\date{\today}

\begin{document}

\begin{abstract}
Higher OH concentration is identified in 
tetragonal barium titanate (BaTiO$_3$) nanorods synthesized by 
a hydrothermal method with 10 vol\% ethylene glycol solvent 
[Inada, M. {\it et al}. {\it Ceram. Int.} {\bf 2015}, {\it 41}, 5581-5587]. 
This is apparently inconsistent with the known fact that 
higher OH concentration in the conventional hydrothermal synthesis 
makes pseudo-cubic BaTiO$_3$ nanocrystals more stable 
than the tetragonal one. 
To understand where and how the introduced OH anions
are located and behave in the 
nanocrystals, we applied {\it ab initio} analysis 
to several possible microscopic 
geometries of OH locations, confirming  
the relative stability of the tetragonal 
distortion over the pseudo-cubic one 
due to the preference of trans-type 
configurations of OH anions. 
We also performed FTIR and XRD analysis, 
all being in consistent with the microscopic 
picture established by the {\it ab initio} 
geometrical optimizations. 
\end{abstract}

\section{Introduction}
\label{intro}
The discovery of a classical ferroelectric barium titanate (BaTiO$_3$) 
dates back to the 1940s~\cite{1942WAI,1943WAI,1947OGA,1945WUL,1987CRO}.
It has a tetragonal structure with a space group symmetry of 
$P4mm$ at room temperature.  
Much attention has been paid to this material because of a variety of 
technical applications ranging from condensers to 
positive temperature coefficient (PTC) 
thermistors~\cite{1959SAB,1961SAU,1975BUR,1981SAK}.  
For example, due to its high dielectric constant, 
BaTiO$_3$ is used as a dielectric layer of multilayer 
ceramic capacitors~\cite{1975BUR,1981SAK}.

\vspace{2mm}
Hydrothermal synthesis~\cite{1999PIN} is one of the widely used 
methods to obtain fine particles of BaTiO$_3$.~\cite{2015INA}
This method enables us to obtain highly pure fine powder with 
small particle size distribution and fairly 
stoichiometric composition~\cite{2010HAY}. 
In the conventional method, however, 
it is known that there remains difficulty in controlling 
the crystal orientation ($c$-axis in the tetragonal structure) 
due to the particle size effect~\cite{2008HOS} and 
the presence of OH groups in the shell region 
(surface and outer layer of crystal)~\cite{2008HOS,2006PET}, 
yielding pseudo-cubic nanocrystals~\cite{2006YAN}
rather than the tetragonal ones. 
Extra thermal annealing at around 1,300 ${}^\circ$C is known 
to be required to recover the tetragonal structure~\cite{2008SAS}. 

\vspace{2mm}
Recently, Inada {\it et al.} have reported that a new hydrothermal 
scheme using 10 vol\% ethylene glycol (EG)~\cite{2015INA}
directly produces tetragonal nanocrystals ($c/a$ = 1.013)
without any extra procedures.
In the present work, we performed the 
thermal gravimetric analysis applied to the samples
and then identified higher OH concentration inside 10 vol\%-EG sample
than 0 vol\%-EG one.
Doubt immediately rises over our finding
because in the conventional hydrothermal synthesis, 
the OH inside BaTiO$_3$ nanocrystals has been widely regarded 
as having a role in stabilizing pseudo-cubic structure.~\cite{2008HOS}
We therefore performed an {\it ab initio} lattice relaxation analysis
to investigate how the OH substitution stabilize tetragonal compared with pseudo-cubic.

\vspace{2mm}
We have found that the geometrical transition
between {\it cis-} and {\it trans-}coordinations
of substituted OH groups depends on their concentrations.
The transition modifies Coulombic interactions of the OH substitutions with 
neighboring Ti cations as well as those with Ba vacancies 
introduced to compensate the charge neutrality. 
The relaxations due to the above modification
can explain the trend of simultaneous contractions along $a$- and $b$-axes
as well as elongations along $c$-axis
when tetragonal structure is stabilized.

\section{Experimental section}
\label{exp}
\subsection{Synthesis}
\label{expsyn}
BaTiO$_3$ nanocrystals were obtained by a hydrothermal synthesis using
Ba-Ti hydroxide precipitation as a precursor.~\cite{2015INA}
In the following, we provide minimum descriptions about 
our synthesis required for understanding further 
characterizations and theoretical analysis. 
Detailed information is given in our preceding paper~\cite{2015INA}: 
The concentration of Ba and Ti in the starting solution was
adjusted to be 0.3 M (mol/L) and 0.2 M, respectively.
First, 1 M BaCl$_2$ (15 mL) and 2 M TiCl$_4$ (5 mL) solutions were
mixed at room temperature, followed by adding 10 M NaOH aqueous solution (10 mL)
so as to precipitate Ba-Ti hydroxide.
In order to obtain 50 mL of starting slurry, 20 mL of deionized (DI) water
or 5 mL (10 vol\%) of EG and 15 mL of DI water were added to the mixture.
The starting slurry was placed into a 100 mL of Teflon-lined stainless steel
autoclave and heated in an oven at 200 ${}^\circ$C for 24 h,
followed by cooling down in ice water.
The products were separated, washed by decantation process repeatedly,
and then dried at 60 ${}^\circ$C overnight.
Hereafter, the samples synthesized by the hydrothermal method with
and without EG are denoted as EG-10 and EG-0, respectively.
Actually, the nucleation and crystal growth of 
BaTiO$_3$ occurred via dissolution-reprecipitation. 
In our previous study~\cite{2015INA}, we reported that the 
nucleation was suppressed by the addition of EG. 
The formation mechanism is also described in the 
literature~\cite{2015INA}.

\subsection{Characterization}
\label{expchar}
The powder X-ray diffraction (XRD) patterns were collected
by X-ray diffractometer (Bruker AXS) with Cu K$\alpha$
radiation (40 kV, 40 mA) at room temperature.
The lattice parameters were refined by
a whole powder pattern decomposition (WPPD) method
assuming a $P4mm$ structural model.
The thermal behavior of the products was evaluated by
thermogravimetric analysis (TGA).
The OH groups were characterized by Fourier transform infrared (FTIR) spectroscopy.

\subsection{Modeling and {\it Ab initio} calculations}
\label{expmodel}
\begin{figure}[htbp]
  \centering
  \includegraphics[width=1.0\hsize]{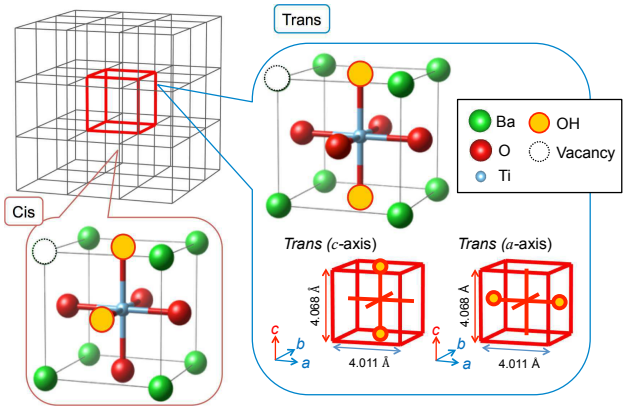}
  \caption{
    Modelling of 7\% OH substitution with 
    {\it cis}- and {\it trans}-coordinations
    by the 3$\times$3$\times$3 supercell. 
    For the latter, we considered the 
    substitutions along $a$- and $c$-axis. 
    Lattice constants shown in the figure 
    are fixed at the experimental values 
    obtained by X-ray diffractions (XRD).
  }
  \label{fig.333}
\end{figure}
\begin{figure}[htbp]
  \centering
  \includegraphics[width=1.0\hsize]{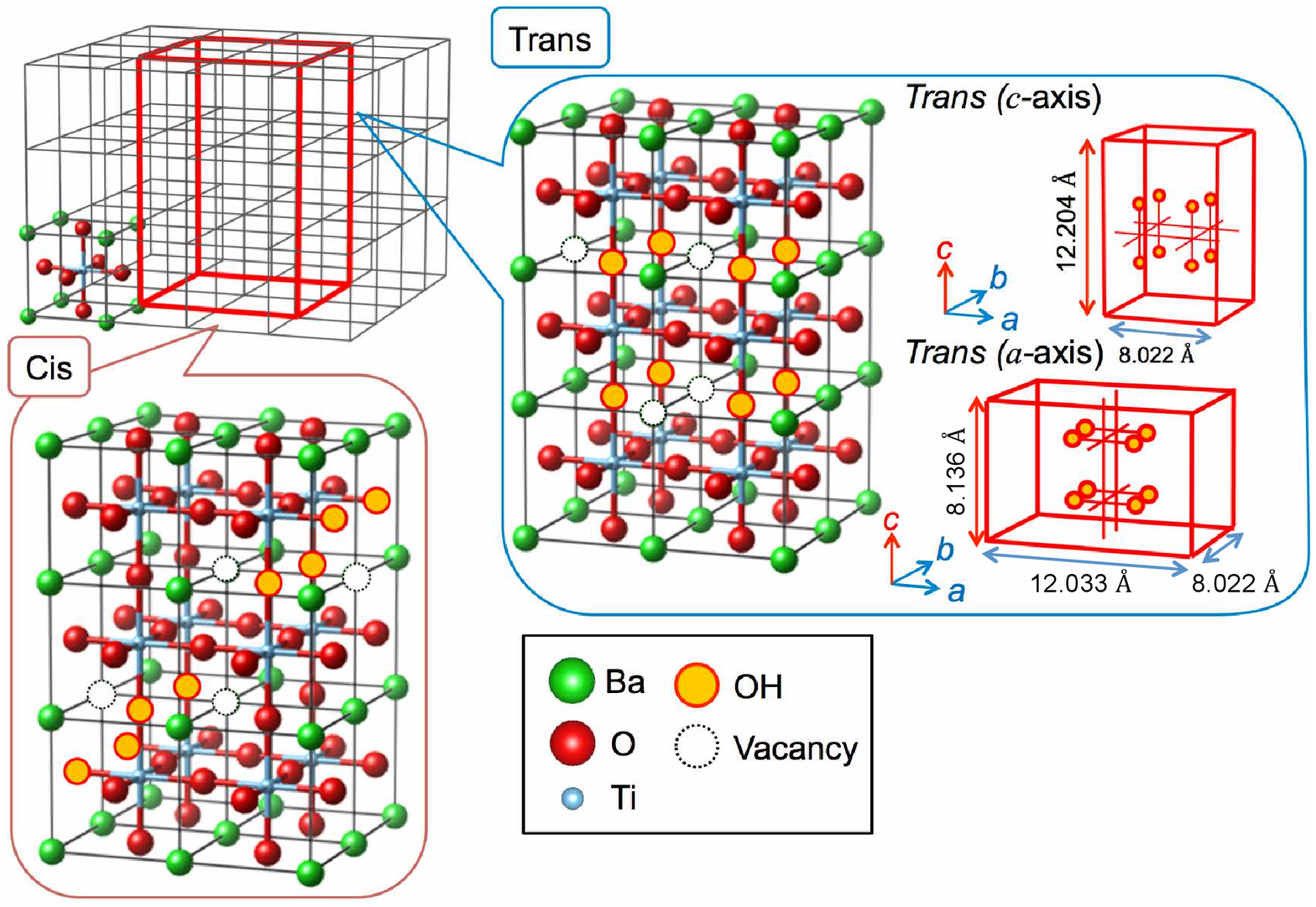}
  \caption{
    Modelling of 17\% OH substitution with 
    {\it cis}- and {\it trans}-coordinations
    by the 4$\times$4$\times$3 supercell. 
    For the latter, we considered the 
    substitutions along $a$- and $c$-axis. 
    Lattice constants shown in the figure 
    are fixed at the experimental values 
    obtained by X-ray diffractions (XRD).
  }
  \label{fig.443}
\end{figure}
\begin{figure}[htbp]
  \centering
  \includegraphics[width=1.0\hsize]{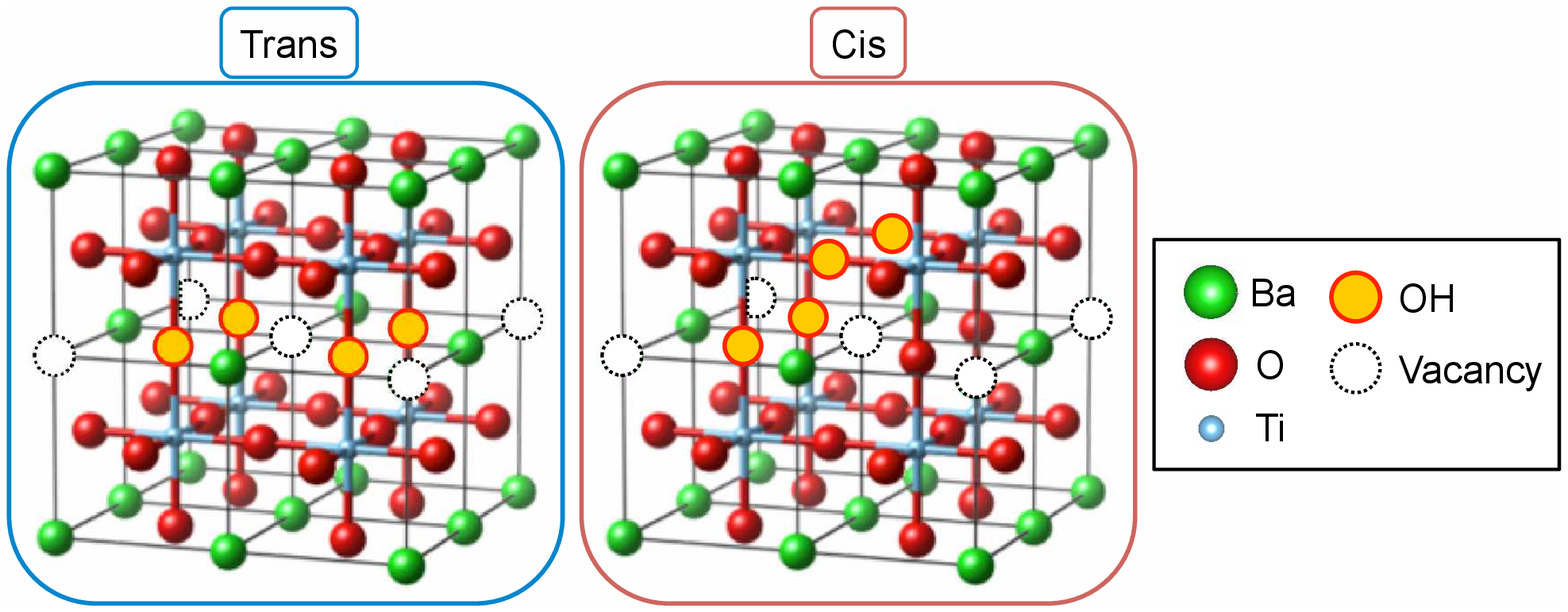}
  \caption{
    Modelling of 50\% OH substitution with 
    {\it cis}- and {\it trans}-coordinations
    by the 2$\times$2$\times$2 supercell. 
    For the latter, we considered the 
    substitutions along $c$-axis. 
  }
  \label{fig.222}
\end{figure}
\begin{figure}[htbp]
  \centering
  \includegraphics[width=1.0\hsize]{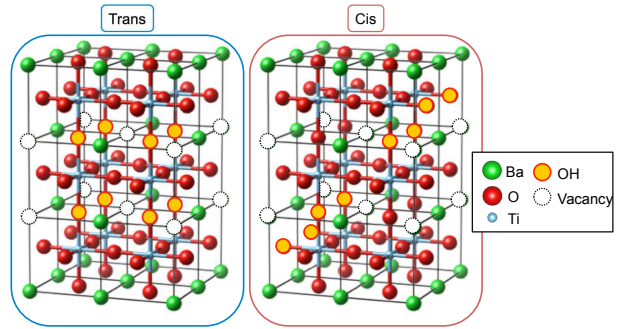}
  \caption{
    Modelling of 67\% OH substitution with 
    {\it cis}- and {\it trans}-coordinations
    by the 2$\times$2$\times$3 supercell. 
    For the latter, we considered the  
    substitutions along $c$-axis. 
  }
  \label{fig.223}
\end{figure}

The present study aims to investigate the stabilization
mechanism of the tetragonal structure depending on the OH concentrations. 
To evaluate relative stability of Ba$_{1-(1/2)x}$TiO$_{3-x}$(OH)$_x$ compounds
with different amounts of substitutions $x$,
we performed {\it ab initio} DFT simulations with the GGA/PBE~\cite{1996PER}
exchange-correlation functional
that has been well tested for both cubic and tetragonal
phases of BaTiO$_3$.~\cite{2008BIL}
The amount of OH in synthesized BaTiO$_3$ 
was estimated from TGA. 
The experimental results strongly indicated 
only upto two substitutions, namely, 
$cis$- and $trans$-coordinations, 
as the possibilities we consider hereafter.
The $c/a$ value, calculated from XRD, 
of as-synthesized BaTiO$_3$ with EG was bigger 
than that of pure tetragonal BaTiO$_3$. 
And also, as-synthesized BaTiO$_3$ included 
a larger amount of OH than pure tetragonal. 
These experimental results indicate that the 
expand of $a$ axis, that means the incorporation 
of OH along $a$-axis, is ``not in accordance'' 
with the experimental observation that
the lattice constants of Ba$_{1-(1/2)x}$TiO$_{3-x}$(OH)$_x$
elongate along $c$-axis compared with those of BaTiO$_3$.
Thus, we compared only three possibilities 
of the substitutions,
{\it i.e.}, OH located at the nearest neighbor [{\it cis}],
or diagonal-wise along $a$-axis ($c$-axis) [{\it trans}-$a$ ({\it trans}-$c$)]
(see Fig.~\ref{fig.333}).
On account of computability, we considered only four stoichiometric
compounds of Ba$_{1-(1/2)x}$TiO$_{3-x}$(OH)$_x$
($x$ = 0.07/0.17/0.50/0.67) within moderate sized simulation
cells in which the numbers of O sites replaced by OH become integers.
They were respectively modeled by 3$\times$3$\times$3, 4$\times$4$\times$3,
2$\times$2$\times$2, and 2$\times$2$\times$3 supercells, {\it i.e.},
by duplicating the original BaTiO$_3$ unit cell, 
as shonw in Figures~\ref{fig.333}, \ref{fig.443},
\ref{fig.222}, and \ref{fig.223}, respectively.
To maintain the charge neutrality, Ba vacancies have to be introduced
(all the possible patterns of locations are provided in
Supporting Information).
For the compensation of the neutrality, 
the reduction of a Ti$^{4+}$ would be 
another possibility to realize the 
counter charge for the OH substitution. 
If that were the case, however, the powder color 
would get blueish, which is not observed in 
our synthesis. 
Moreover in general, Ba vacancies are easily 
introduced in the BaTiO$_3$ particles when synthesized 
by precipitation method because Ba ion is stable in 
alkaline condition of the reacting solution. 
These are the reason why we use the model of 
Ba vacancy and OH substitution.
The model structures described above were generated by
BIOVIA Materials Studio Visualizer~\cite{2017MSV}, 
and then the atomic positions
within the simulation cells were further relaxed under the
fixed lattice parameters (set as experimental value).

\vspace{2mm}
The geometry optimizations were carried out using the BFGS algorithm
implemented in CASTEP~\cite{2005CLA} with thresholds for the energy
convergence (1.0$\times 10^{-3}$ eV/cell), the force convergence
(0.05 eV/\AA), the stress convergence (0.1 GPa),
and the displacement convergence (0.002 \AA).
Monkhorst-Pack $k$-point meshes of 1$\times$1$\times$1,
1$\times$1$\times$1, 2$\times$2$\times$2, and 2$\times$2$\times$1
were adopted for the supercells with $x$ = 0.07, 0.17, 0.15, and 0.67,
respectively, each of which corresponds to $k$-point separation of 0.08 \AA$^{-1}$.
Ionic cores were described by the on-the-fly ultrasoft pseudo potentials
implemented in CASTEP~\cite{2005CLA}: [He] for O, [Ne] for Ti, and [Kr]4$d^{10}$ for Ba.
Kohn-Sham orbitals were expanded in terms of planewaves with $E_{cut}$ = 340 eV
that is determined by the criterion that the total energy converges
within 1.0$\times 10^{-4}$ eV/cell.

\section{Results and Discussion}
\label{results}
\subsection{Crystal structure of the BaTiO$_3$ nanocrystals and OH in those crystals.}
\label{struct}
\begin{figure}[htbp]
  \centering
  \includegraphics[width=1.1\hsize]{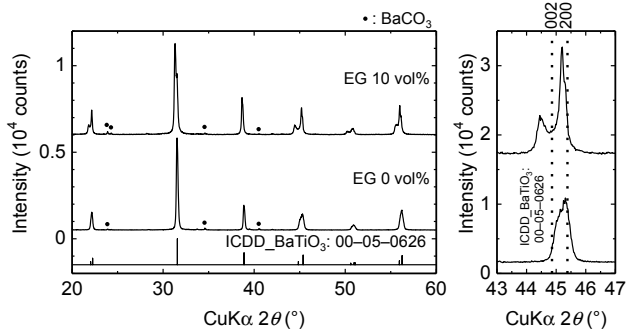}
  \caption{
    X-ray diffraction (XRD) patterns of
    BaTiO$_3$ synthesized with and without ethylene glycol 
    (EG-10 and EG-0, respectively).
    (a) Whole area and (b) selected area that indicates reflections from (200) planes.
    Dotted lines show peak positions of (200) and (002) directions
    from ICDD: 00-005-0626.
  }
  \label{fig.xRay}
\end{figure}

\begin{figure}[htbp]
  \centering
  \includegraphics[width=1.0\hsize]{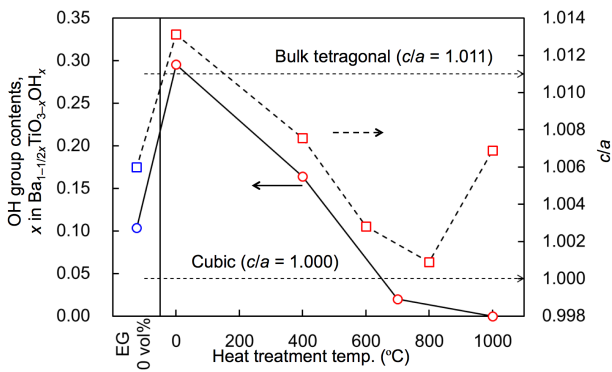}
  \caption{
    Changes in OH contents and $c/a$ of BaTiO$_3$ synthesized
    with 10 vol\% EG versus heat treatment temperature.
    Blue plots indicate the results without EG.
  }
  \label{fig.ohContents}
\end{figure}

Fig.~\ref{fig.xRay} shows XRD patterns of the EG-10 and EG-0 samples.  
It was confirmed that the samples consist of a major phase of BaTiO$_3$
and a trace impurity of BaCO$_3$ as shown in Fig.~\ref{fig.xRay} (a).
The BaCO$_3$ impurity was formed by the reaction between Ba$^{2+}$ and CO$_3^{2-}$
dissolved from the atmosphere.
The magnified view of the (200) and (002) reflections shown in Fig.~\ref{fig.xRay} (b)
reveals that these reflection peaks are clearly separated for the EG-10 sample
while the one broad peak is observed for the EG-0 sample.
This means that the EG-10 and EG-0 samples exhibit high and low tetragonality, respectively.
It was also found that the diffraction angles of the EG-10 sample are lower than
those of the EG-0 sample and bulk BaTiO$_3$ (dotted lines),
indicating lattice expansion due to the incorporation of the
OH groups in the crystals.
Fig.~\ref{fig.ohContents} plots the content of OH groups
determined by TGA along with the $c/a$
ratio as a function of heat-treatment temperature.
The content of OH groups in the EG-10 sample is about three times
as many as that in the EG-0 sample.
The EG-10 sample has a higher $c/a$ ratio than that of bulk BaTiO$_3$,
indicating the enhancement of tetragonality due to the incorporation of the OH groups.
The OH content and the $c/a$ ratio decrease with increasing heat-treatment
temperature up to 800 ${}^\circ$C, 
suggesting a significant correlation between the content of
OH groups and the $c/a$ ratio of the BaTiO$_3$ crystals. 

\subsection{Characterization of the OH groups in the crystals by FTIR and DFT calculations}
\label{char}
\begin{figure}[htbp]
  \centering
  \includegraphics[width=1.1\hsize]{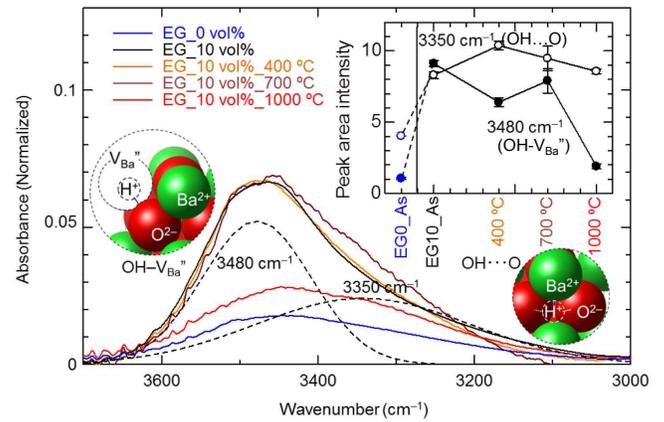}
  \caption{
    FTIR spactra of the OH groups in the BaTiO$_3$ nanocrystals
    synthesized with 10 vol\% EG before and after heat treatment at 400-1000 ${}^\circ$C.
    For comparison, the result without EG is also shown in blue line. 
    The peak area intensities versus the heat treatment temperatures are inset.
  }
  \label{fig.ftir}
\end{figure}

FTIR~(Fourier transform infrared spectrometer)
analysis was carried out in order to characterize the OH groups in the crystals.
The result is shown in Fig.~\ref{fig.ftir}.
We found a peak around at 3500 cm$^{-1}$, which can be deconvoluted with a slightly
sharp peak around at 3480 cm$^{-1}$ and a broad peak around at 3350 cm$^{-1}$,
indicating that there are two kinds of OH groups in the crystals.
The peak area intensity of the EG-10 sample is larger than that of 
the EG-0 sample; in particular, the area intensity of the peak around at
3480 cm$^{-1}$ is much larger for the EG-10 sample than for the EG-0 sample.
As explained later, our DFT gives 
two different relaxed positions of substituted OH, 
which can be identified to each of peaks, namely, 
3480 cm$^{-1}$ being attributed to OH-V$_\mathrm{Ba}$ 
defect complexes, and 3350 cm$^{-1}$ to OH...O hydrogen bonds.
The area intensity of a peak around at 3480 cm$^{-1}$
decreases by heat treatment as shown in the inset of Fig.~\ref{fig.ftir}.

\subsection{{\it Ab initio} Analysis}
\label{first}
\begin{figure}[htbp]
  \centering
  \includegraphics[width=1.0\hsize]{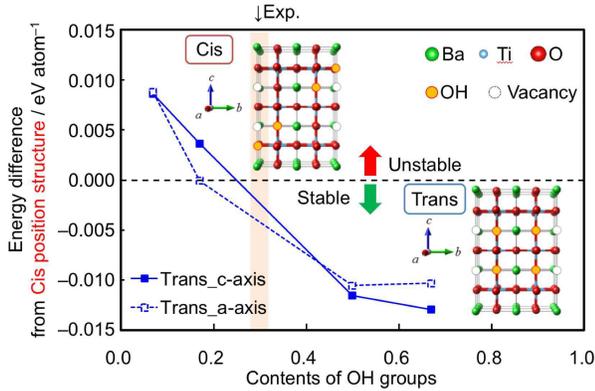}
  \caption{
The total energies of the {\it trans}-coordinations along $a$- and $c$-axes 
with respect to that of the {\it cis}-coordinations. 
The region above the broken line indicates the {\it cis}-coordination is stable. 
The region bellow the broken line indicates the {\it trans}-coordination is stable.
  }
  \label{fig.cisTrans}
\end{figure}

{\it Ab initio} analysis was carried out to
gain microscopic insights into how the substituted OH-sites
play a role in stabilizing the tetragonal crystal structure. 
As explained in the previous section, 
we prepared four different supercells to model 
various substitution, $x$. 
We compared the energies of 
{\it cis}- and {\it trans}-coordinations 
of OH-substituents as a function of $x$ 
as shown in Fig.~\ref{fig.cisTrans}. 
We see that the relative stability inverts 
as the substituent concentration $x$ increases. 

\begin{figure}[htbp]
  \centering
  \includegraphics[width=1.1\hsize]{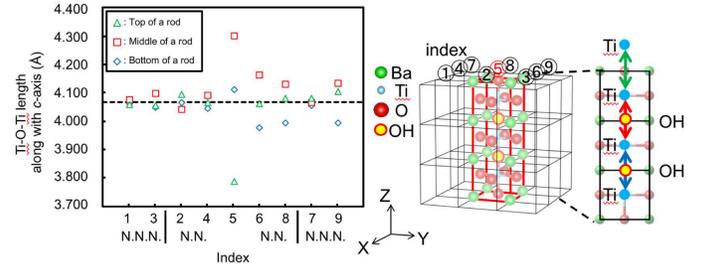}
  \caption{
    Spatial relaxations of the elongated bonding lengths 
    along $c$-axis, evaluated at $x$=0.07 
    described by 3$\times$3$\times$3 supercell. 
    Right panel gives the convention of indexing for the 
    'rods', O-Ti$^{(top)}$-O-Ti$^{(center)}$-O-Ti$^{(bottom)}$-O, 
    along $c$-axis. 
    Upper and lower sides of Ti$^{(center)}$ at 'rod\#5' are 
    substituted into OH ({\it trans}-coordination). 
    Lengths of O-Ti$^{(top)}$-O, O-Ti$^{(center)}$-O, and 
    O-Ti$^{(bottom)}$-O are plotted in the left panel, 
    showing the local distortion at the substituted site 
    is relaxed down over the nearest neighbors (N.N.) and 
    the next nearest neighbors (N.N.N.).
  }
  \label{fig.distSmall}
\end{figure}

\begin{figure}[htbp]
  \centering
  \includegraphics[width=1.1\hsize]{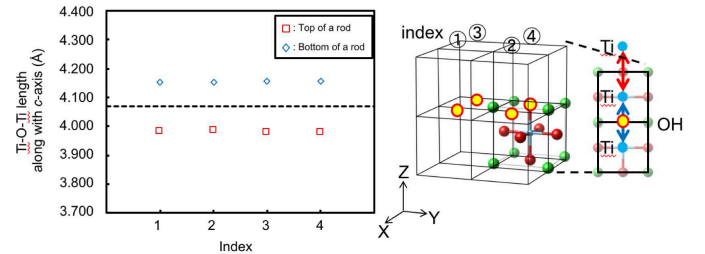}
  \caption{
    For the larger $x$ (OH concentration), 
    the distortions due to the OH substitutions 
    (changes in the bonding lengths along $c$-axis) 
    are cooperatively occurs over the crystal. 
    Right panel gives the convention of indexing for the 
    'rods', O-Ti$^{(top)}$-OH-Ti$^{(bottom)}$-O, 
    along $c$-axis. 
    The bond lengths are evaluated at $x=0.5$ 
    described by 2$\times$2$\times$2 supercell, 
    plotted in the left panel over the rods \#1$\sim$\#4.}
  \label{fig.distLarge}
\end{figure}

\vspace{2mm}
The fact that the {\it trans}-coordination gets
more stable for larger $x$ can be roughly connected with 
the {\it physical pictures}  
of the distortion caused by the OH-substitution.
Figures~\ref{fig.distSmall} and~\ref{fig.distLarge} show 
how the elongations of O-Ti-O bond lengths appear
around the OH-sites with different concentrations, 
$x=0.07$ and $x=0.5$, respectively. 
While the distortion occurs locally around the OH-sites 
for the smaller $x$ (Fig.~\ref{fig.distSmall}), 
it spreads over the whole system uniformly 
for the larger $x$ (Fig.~\ref{fig.distLarge}). 
The latter has a large enough number of the OH-sites
to get the distortions cooperatively accommodated within the system,
which makes energy loss less than the former
where the local distortion causes the energetically unstable state. 

\begin{figure}[htbp]
  \centering
  \includegraphics[width=1.0\hsize]{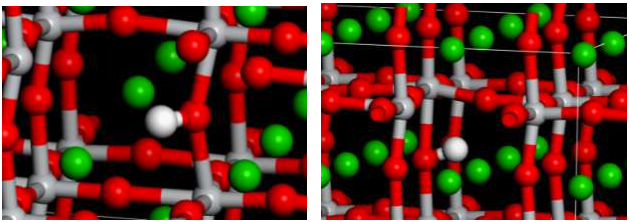}
  \caption{
    Local relaxations of atomic positions around 
    the substituted OH, evaluated at $x=0.07$ 
    described by 3$\times$3$\times$3 supercell.
    White (red) balls represent H (O) atoms. 
    Two cases with and without a Ba vacancy are 
    shown in left and right panels, respectively. 
    In the former case, the OH is found to be 
    attracted toward the vacancy due to 
    its positively charged background. 
    In the latter case instead, the OH is 
    attracted toward the neighboring O to form 
    hydrogen bondings.}
  \label{fig.BaVac}
\end{figure}

\vspace{2mm}
Fig.~\ref{fig.BaVac} shows the lattice relaxation 
around the substituted OH-site, evaluated by 
3$\times$3$\times$3 supercell with $x=0.07$. 
From Figures~\ref{fig.distSmall} and~\ref{fig.distLarge}, 
we could see two possible effects of causing distortions by the substituents, 
'(a) OH-site solely gets attracted by the vacancy V$_{\rm Ba}$' 
and 
'(b) the {\it trans} location of two OH along $c$-axis 
gets the OH-Ti-OH bond elongated along the axis'.
The effect (a) could be explained by the fact 
that the vacancy would be charged negatively 
if the positive Ba$^{2+}$ ion was not there, 
attracting a proton of a OH ion via 
electrostatic interactions. 
Comparing the relaxations of OH positions with and without 
V$_{\rm Ba}$ nearby (left and right panels of Fig.~\ref{fig.BaVac}), 
we can see the formation of the hydrogen bonding between 
OH and nearest-neighboring O-site when without vacancy. 
This contrast would explain two different absorption peaks 
in FTIR (Fig.~\ref{fig.ftir}) 
at 3480 cm$^{-1}$ and 3350 cm$^{-1}$.
The latter peak would be identified to 
the lattice vibrations with weakened coupling 
due to the formation of the hydrogen bonding.

\vspace{2mm}
The effect (b), the elongation of OH-Ti-OH bond, 
could be qualitatively understood 
as a consequence of the weakened electrostatic interaction with Ti 
caused by replacing O$^{2-}$ with OH$^{-}$.
The displacement of Ti is, however, found to be 
small being around 1.049$\sim$1.060 in terms of the ratio 
of the longer Ti-O bond length to the shorter one 
(Fig.~\ref{fig.distLarge}). 
The smaller displacement would be explained as follows: 
Since the larger displacement of Ti from its original center would yield 
the two larger dipole moments in the opposite direction, 
the resultant dipole-dipole interactions could give rise to an increase in the system energy.
Thus, the possible displacement remains within the smaller amount
so as to prevent the system from increasing the energy.
\begin{figure}[htbp]
  \centering
  \includegraphics[width=0.8\hsize]{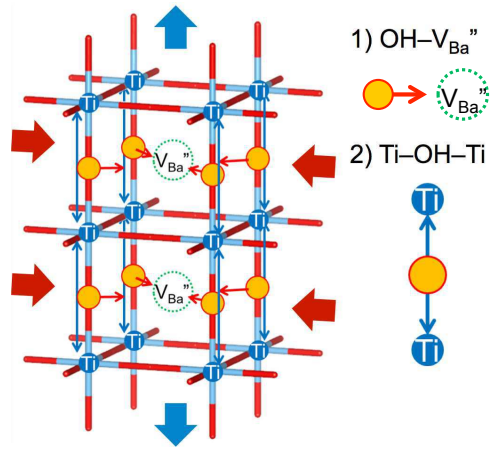}
  \caption{
    A schematic picture showing how the substituted OH 
    stabilizes the tetragonal structure with higher concentration. 
    They are attracted toward a Ba vacancy, which 
    emerges to realize the charge neutrality, yielding 
    contractions of $a$ and $b$ bond lengths. 
    On top of that, the reduction in the negative stoichiometric charge     
    (O$^{2-}\to$ OH$^{-}$) brings about the elongation of 
    bonding length along $c$-axis.}
  \label{fig.schem}
\end{figure}

\vspace{2mm}
The above effects, 
'(a) attraction of OH by V$_{\rm Ba}$' and
'(b) elongation of OH-Ti-OH bonds', 
would explain the trend in stability of
the tetragonal strucutre relative to the pseudo-cubic one 
depending on the OH concentration, $x$. 
For the region where the {\it cis}-coordination 
is stabilized, {\it e.g.}, $x=0.07$, 
the OH-sites get attracted toward the Ba-vacancies 
individually and uniformly, giving rise to the pseudo-cubic structure. 
For the larger $x$, {\it e.g.}, $x=0.25$, where 
the {\it trans}-coordination is stabilized instead, 
the effect (a) promotes the contraction of $a$- and $b$-axe 
(see Fig.~\ref{fig.schem}), 
while the effect (b) yields the elongation along $c$-axis, 
resulting in the stabilization of the tetragonal structure. 
Thermal gravimetric analysis (TGA) applied to the present 
EG-synthesized samples estimates $x \sim 0.28$, 
dropping at the region where the tetragonal structure 
gets stabilized, being consistent with the experimental fact. 

\vspace{2mm}
The above scenario is not contradicting to the fact that 
the conventional hydrothermal synthesis gives  
pseudo-cubic nanocrystals even with higher $x$ 
as long as we regard it as the 'core-shell' structure~\cite{2008HOS} 
where the doped OHs are not incorporated into 
the 'core' region but located only 
at the 'shell' (surface of the core) randomly.
The present analysis using supercells corresponds to 
the assumption that the doped OHs are located uniformly 
inside the entire crystal, being the case only 
for the EG-synthesized sample. 

\section{Conclusion}
\label{conc}
We investigated the role of substituted OH 
in stabilizing tetragonal anisotropy of 
the BaTiO$_3$ nanocrystal synthesized 
by a hydrothermal scheme using ethylene glycol (EG). 
Thermogravimetric analysis was carried out to identify 
the concentration of OH, revealing that the EG-synthesized sample
has around three times larger amount of OH concentration
compared to the sample systhesized by the conventional 
hydrothermal scheme. 
Since in the conventional scheme the introduced OH 
has been regarded as playing a role in stabilizing
the pseudo-cubic isotropic structure, 
the present tetragonal anisotropic structure
involving the large amount of OH concentration
stimulated curiosity.

\vspace{2mm}
The apparent contradiction is finally attributed to 
how the introduced OH groups are distributed throughout a sample.
Employing structural models with uniform OH 
distributions, our {\it ab initio} geometrical 
relaxation analysis concluded 
the stabilization of tetragonal anisotropy 
in the higher OH concentration range where 
the {\it trans}-coordinations of substitutions 
were preferred. 
In pseudo-cubic nanocrystals synthesized 
by the conventional scheme, instead, the introduced OH 
can be thought of as being distributed only within the 'shell' region
over the 'core' region, not uniformly, and hence
the present analysis is beyond the scope of the core-shell interpretation.

\vspace{2mm}
The predicted stabilization was explained 
by several mechanisms such as, 
the cooperative accommodation of the lattice deformations, 
electrostatic interactions between Ba vacancies and OH, 
and the elongation along the {\it trans}-coordination of OHs
due to less attraction between anion site and Ti 
caused by the reduction in negative charge from O$^{2-}$ to OH$^{-}$.

\section{Author information}
Competing financial interests: The authors declare no competing financial interests.

\section{Supporting Information}
\label{SI}
All the possible patterns of Ba vacancies and OH anions in 
Ba$_{1-(1/2)x}$TiO$_{3-x}$(OH)$_{x}$ ($x = 0.07/0.17/0.50/0.67$)
are provided, which is mentioned in ``Modeling and {\it Ab initio} calculations''.
This information is available free of charge via the Internet at http://pubs.acs.org.

\section{Acknowledgement}  
This work was supported by the Grant-in-Aid for Scientific
Research on Innovative Areas ``Mixed Anion'' project (JP16H06439,
JP16H06440 and 17H05478) from MEXT.  
All the computation in this work has been performed
using the facilities at Research Center for Advanced
Computing Infrastructure in JAIST.
The FTIR analysis was carried out using FTIR620
at the Center of Advanced Instrumental Analysis, Kyushu University.
K. Hongo is also grateful for financial support from a
KAKENHI grant (17K17762), PRESTO (JPMJPR16NA) and the Materials research
by Information Integration Initiative (MI$^2$I) project of the
Support Program for Starting Up Innovation Hub from Japan Science
and Technology Agency (JST).
R.M. is also grateful for financial supports
from MEXT-KAKENHI (16KK0097), 
from FLAGSHIP2020 (project nos. hp170269 and hp180175 at K-computer), 
from Toyota Motor Corporation, from I-O DATA Foundation, 
and from the Air Force Office of Scientific Research 
(AFOSR-AOARD/FA2386-17-1-4049). 
K.Hayashi is supported by Elements Strategy Initiative
to Form Core Research Center, MEXT, Japan.

\bibliography{references}

\end{document}